\documentclass[reprint,aps,prl,superscriptaddress,floatfix]{revtex4-2}

\usepackage{amsmath}
\usepackage{amssymb}
\usepackage{graphicx}
\usepackage{bm}

\begin{document}

\title{Joint Amplitude--Phase Optimization of Broadband Lasers Raises Absolute Two-Plasmon-Decay Thresholds Above Coherence-Time Predictions}

\author{A. S. Joglekar}
\affiliation{Ergodic LLC, Seattle, WA 98103, USA}
\affiliation{Laboratory for Laser Energetics, University of Rochester, Rochester, NY 14623, USA}
\author{R. K. Follett}
\affiliation{Laboratory for Laser Energetics, University of Rochester, Rochester, NY 14623, USA}
\author{D. H. Froula}
\affiliation{Laboratory for Laser Energetics, University of Rochester, Rochester, NY 14623, USA}
\author{J. P. Palastro}
\affiliation{Laboratory for Laser Energetics, University of Rochester, Rochester, NY 14623, USA}

\date{\today}

\begin{abstract}
Broadband lasers suppress two-plasmon decay, a source of fuel preheat in inertial confinement fusion.
Coherence-time predictions depend on the power spectrum but are insensitive to spectral phase.
Differentiable wave simulations show that phase optimization raises the absolute instability threshold at fixed power spectrum.
At 1\% bandwidth, joint amplitude--phase optimization doubles the median random-phase threshold.
A growth-rate budget reveals reduced available coupling and pump--daughter alignment.
The advantage persists over 1--4\% bandwidth and at OMEGA- and ignition-scale conditions.
\end{abstract}

\maketitle

Parametric instabilities driven by laser--plasma interactions remain a central challenge for inertial confinement fusion (ICF)~\cite{Craxton2015}.
In direct-drive ICF, two-plasmon decay (TPD), in which a laser photon decays into a pair of electron plasma waves (EPWs) near the quarter-critical density surface, generates suprathermal electrons that preheat the fuel and degrade implosion performance~\cite{Kruer1988,yan_generating_2012}.
Broadband lasers suppress these instabilities. Bandwidth decorrelates the parametric drive and raises the absolute instability threshold~\cite{Follett2019, follett_thresholds_2021, wen_suppressing_2021, bates_suppressing_2023}, and recent advances in broadband ultraviolet lasers have renewed interest in this approach~\cite{dorrer_broadband_2021,froula_future_2025}.
The smoothing schemes fielded to date already impose particular spectral phases. Smoothing by spectral dispersion uses a deterministic frequency-modulation sweep~\cite{skupsky_improved_1989}, while induced spatial incoherence uses random phases~\cite{lehmberg_use_1983}. These phases were chosen to smooth hydrodynamic imprint, not to control parametric growth.

Existing numerical results express that suppression through a coherence-time scaling. Thresholds for broadband drives with random spectral phases collapse onto a curve set by the field coherence time $\tau_c$~\cite{Follett2019}.
By the Wiener--Khinchin theorem, however, $\tau_c$ is a functional of the power spectrum alone and cannot distinguish two drives that share a spectrum but differ in spectral phase.
Here we treat both the spectral amplitudes $A_j$ and phases $\varphi_j$ of an $N$-line drive as design variables and search the joint space, in which each existing smoothing scheme is a single point, by gradient descent through a differentiable enveloped wave solver~\cite{joglekar_adept_2023,jax_github_2018,joglekar_gnr_2024}.
At fixed power spectrum, optimizing phase raises the threshold from the random-phase median of $2.8\,I_\mathrm{mono}$ to $4.1\,I_\mathrm{mono}$, where $I_\mathrm{mono}$ is the threshold for a monochromatic drive \cite{simon_inhomogeneous_1983}, but the resulting transform-limited (TL) pulse train has peak-to-average intensity ratio $R=32$.
Allowing amplitudes and phases to vary together while penalizing R yields $5.6\,I_\mathrm{mono}$ at $R=3.6$, above the best tested TL line-count result of $4.5\,I_\mathrm{mono}$ at $R=16$. Less constrained joint designs reach $6.4\,I_\mathrm{mono}$.

\begin{figure}[!t]
    \includegraphics[width=\columnwidth]{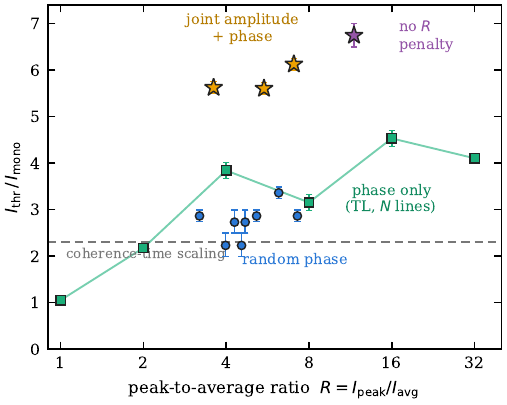}
    \caption{\label{fig:headline}
    Absolute TPD threshold vs.\ peak-to-average ratio $R = I_\mathrm{peak}/I_\mathrm{avg}$ at $1\%$ bandwidth under OMEGA-scale conditions (protocol in End Matter).
    Circles: eight equal-amplitude random-phase drives ($N=32$).
    Squares: the TL line-count benchmark, equal-amplitude transform-limited drives with $N = 1$--$32$ ($R = N$).
    Stars: joint amplitude--phase optima ($N=32$), with a peak-ratio penalty (gold) or without (purple).
    Dashed: the coherence-time threshold $I_c$~\cite{Follett2019}.
    }
\end{figure}

We solve the enveloped wave equations~\cite{myatt_wave-based_2017, myatt_lpse_2019} for the EPW envelope $E_h$ coupled to a prescribed laser envelope $E_0$ at carrier frequency $\omega_0$. For normal incidence on a linear density profile spanning $n_e/n_c \in [0.18, 0.28]$, the governing equation is
\begin{widetext}
\begin{equation}\label{eq:epw}
    \nabla\cdot\!\left[2i\omega_{p0}\!\left(\partial_t + \gamma_h\right) + 3 v_{te}^2\,\nabla\nabla\cdot{} + \omega_{p0}^2 - \omega_{pe}^2\right] E_h = \frac{e\,\omega_{p0}}{2 m_e \omega_0}\,\nabla\cdot\!\left[\nabla(E_h^{*}\cdot E_0) + \left(1 - \tfrac{\omega_0}{\omega_{p0}}\right) E_0\,\nabla\cdot E_h^{*}\right] e^{i(2\omega_{p0}-\omega_0)t},
\end{equation}
\end{widetext}
where $\omega_{p0}$ is the envelope frequency, $\omega_{pe}(x)$ is the local plasma frequency, $v_{te}$ is the electron thermal velocity, and $\gamma_h$ collects Landau and collisional damping.
The drive is a sum of $N$ plane waves,
\begin{equation}\label{eq:E0}
    E_0 = \hat{y} \sum_{j=1}^{N} A_j \, e^{i(k_j x - \delta\omega_j t + \varphi_j)},
\end{equation}
with uniform frequency offsets $\delta\omega_j$ spanning the bandwidth and $\sum_j A_j^2$ fixed by the average intensity.
Reference configurations are random-phase (equal $A_j$, $\varphi_j$ random, as in Ref.~\cite{Follett2019}) and transform-limited ($\varphi_j = 0$, equal $A_j$).
We examine two regimes bracketing present-day and ignition-scale direct-drive conditions, OMEGA-scale ($T_e = 2$~keV, $L_n = 200~\mu$m) and ignition-scale ($T_e = 4$~keV, $L_n = 600~\mu$m), both at a laser wavelength $\lambda_0 = 0.351~\mu$m. Here, $T_e$ is the electron temperature, and $L_n$ is the density scale length.
We identify threshold from the spatially integrated EPW energy over $10$--$40$~ps, combining a duration-normalized exposure with a persistent-growth test; this criterion reproduces the monochromatic and independent-two-beam limits and is tested on $100$~ps trajectories (threshold protocol in End Matter).
The phase-only and joint amplitude--phase optimizations use $N=32$ spectral lines.

Figure \ref{fig:headline} shows the threshold behavior for various designs. At fixed spectrum, the eight $N=32$ random-phase seeds span $2.2$--$3.4\,I_\mathrm{mono}$, with a median of $2.8\,I_\mathrm{mono}$. Optimizing their phases at fixed amplitudes converges to a TL pulse train that reaches $4.1\,I_\mathrm{mono}$, a factor of $1.5$ above the random-phase median without changing the power spectrum. The same average power is then concentrated into periodic peaks with $R=N=32$.

Amplitude freedom can retain the threshold increase without such extreme peak intensities.
As a reference, the TL line-count scan varies the spectrum along with $N$ and therefore forms a benchmark rather than a fixed-spectrum phase test. It peaks at $4.5\,I_\mathrm{mono}$ for $N=16$ ($R=16$).

We obtain the joint amplitude--phase optima (stars in Fig.~\ref{fig:headline}) by minimizing late-time EPW energy over amplitudes and phases, averaged across a band of drive intensities, with a smooth penalty for exceeding a target peak ratio $C_\mathrm{max}$ (optimization in End Matter). Because the penalty is smooth, $C_\mathrm{max}$ labels the optimization target; all performance comparisons use the measured peak ratio $R$.
The optima have nonuniform spectral amplitudes and nonmonotonic phase profiles (Fig.~\ref{fig:spectra}).
The most peak-constrained design reaches $5.6\,I_\mathrm{mono}$ at $R=3.6$, exceeding the best tested TL line-count threshold of $4.5\,I_\mathrm{mono}$ at a $4.4\times$ lower peak ratio.
Relaxing the target raises the joint threshold to $6.4\,I_\mathrm{mono}$ as the measured peak ratio rises to approximately $12$.
Their normalized time-averaged field magnitudes remain near the random-phase values, avoiding the extreme temporal concentration of the TL drive.

\begin{figure}
    \includegraphics[width=\columnwidth]{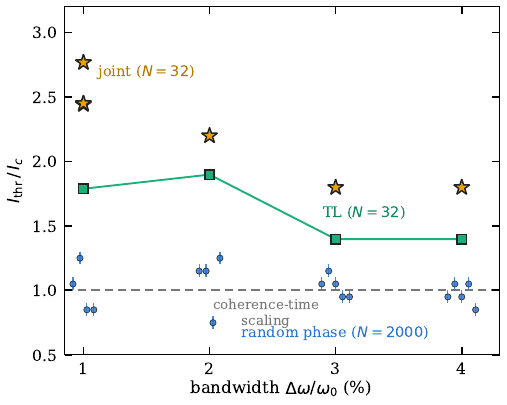}
    \caption{\label{fig:bandwidth}
    Threshold vs.\ bandwidth, normalized to the coherence-time threshold $I_c(\Delta\omega)$~\cite{Follett2019}.
    Circles and bars: selected random-phase realizations and threshold brackets ($N=2000$; sampling in End Matter), offset horizontally for clarity.
    Squares: transform-limited drives; stars: joint optima, both with $N=32$ (target $C_\mathrm{max}=5$ across bandwidth, with additional peak-ratio targets at $1\%$).
    }
\end{figure}

Figure~\ref{fig:bandwidth} normalizes each threshold by the bandwidth-dependent coherence-time prediction $I_c(\Delta\omega)$ rather than by $I_\mathrm{mono}$. The plotted value therefore measures how far each design rises above that prediction.
The $N=2000$ random-phase reference includes thresholds on both sides of the coherence-time prediction at $1$--$4\%$ bandwidth; the optimized and TL drives retain $N=32$.
Re-optimizing at $2\%$, $3\%$, and $4\%$ bandwidth yields joint thresholds of $2.2$, $1.8$, and $1.8\,I_c$, respectively, compared with $2.4\,I_c$ at $1\%$.
At $3$--$4\%$ bandwidth, the TL threshold is $1.4\,I_c$, while the joint optimum remains $20$--$30\%$ above TL and does so at $R\approx3$--$5$ rather than $R=32$.

\begin{figure}
    \includegraphics[width=\columnwidth]{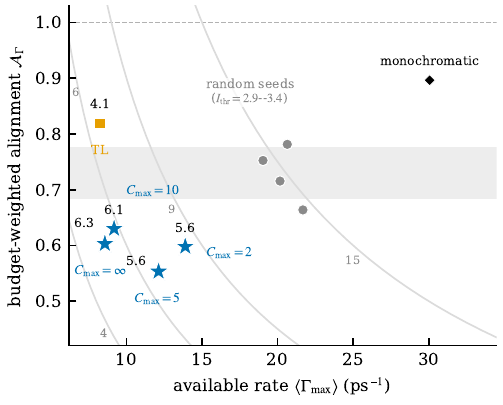}
    \caption{\label{fig:mechanism}
    Pump--field coupling budget [Eq.~\eqref{eq:budget}] at a common average intensity of $4.29\,I_c$.
    Available rate $\langle\Gamma_{\max}\rangle$ vs.\ budget-weighted alignment $\mathcal A_\Gamma$ for four random-phase seeds, the transform-limited comb, four joint optima labeled by peak-ratio target $C_\mathrm{max}$, and a monochromatic reference (black diamond).
    Black numbers: independently measured thresholds in units of $I_\mathrm{mono}$.
    Gray hyperbolas show realized growth $\langle\Gamma_{\mathrm{TPD}}\rangle$ (ps$^{-1}$), and the band marks the random-seed ensemble $\mathcal A_\Gamma = 0.73\pm0.05$ ($n=11$).
    }
\end{figure}

To determine why the joint optimum exceeds the fixed-spectrum phase optimization and the tested TL line-count benchmark, we project the pump source onto the simulated daughter field, which measures the pump--field energy exchange exactly.
Alongside the EPW energy $\mathcal{E}_h(t)$, we record the complex pump-to-EPW power $\mathcal{P}_{\mathrm{TPD}}(t)$ [Eq.~\eqref{eq:Ptpd}]. The quantity $\Gamma_{\mathrm{TPD}} = 2\,\mathrm{Re}\,\mathcal{P}_{\mathrm{TPD}}/\mathcal{E}_h$ is the instantaneous rate at which the pump delivers log-growth to the field that exists, while $\Gamma_{\max} = 2\,|\mathcal{P}_{\mathrm{TPD}}|/\mathcal{E}_h$ is the rate the same pump would deliver at the optimal global phase.
Over an averaging interval, the budget factorizes identically into the two quantities a drive can suppress,
\begin{equation}\label{eq:budget}
    \langle\Gamma_{\mathrm{TPD}}\rangle \;=\; \mathcal A_\Gamma\,\langle\Gamma_{\max}\rangle,
\end{equation}
where $\langle\Gamma_{\max}\rangle$ is the \emph{available} coupling rate and $\mathcal A_\Gamma\in[-1,1]$ is the fraction the pump \emph{realizes} through its alignment with the daughter pair [Eq.~\eqref{eq:alignment}].
At a common average intensity, the realized growth is strongly anticorrelated with the independently measured thresholds across the nine broadband designs ($r=-0.92$, $\rho=-0.86$). It diagnoses pump--field coupling rather than predicting every pairwise threshold difference.

The two-factor plane in Fig.~\ref{fig:mechanism} distinguishes the suppression mechanisms. The monochromatic reference has $\langle\Gamma_{\max}\rangle=30.1~\mathrm{ps}^{-1}$ and $\mathcal A_\Gamma=0.90$. Random-phase drives make $19$--$22~\mathrm{ps}^{-1}$ of growth available and realize $\mathcal A_\Gamma=0.66$--$0.78$ of it.
The TL comb suppresses growth primarily by concentrating the pump into short bursts. It cuts the available rate by $2.4\times$ but realizes \emph{more} of what remains than random phase ($\mathcal A_\Gamma=0.82$). Because the TPD source is linear in $E_0$, we quantify the drive-envelope contribution by the normalized time-averaged magnitude of the launched field, $f_A=\langle|E_0(0,t)|\rangle/\langle|E_0(0,t)|^2\rangle^{1/2}$, averaged over one repetition period. TL has $f_A=0.42$, compared with $0.87$ for random phase. If the normalized pump--daughter overlap is unchanged, this ratio predicts a $2.0\times$ availability reduction, compared with $2.4\times$ measured, and accounts for approximately $81\%$ of the reduction in log-ratio terms.

The joint optima behave differently. Their normalized time-averaged field magnitudes remain near random phase ($f_A=0.80$--$0.88$), yet their available rates are $1.4$--$2.3\times$ lower. The $f_A$ scaling accounts for at most $10\%$ of that reduction in log-ratio terms; most therefore appears as a smaller period-averaged normalized overlap between the realized field and its source, since $\Gamma_{\max}=2|\mathcal{P}_{\mathrm{TPD}}|/\mathcal{E}_h$. This global diagnostic does not identify the spatial or modal origin.
The optima also realize only $\mathcal A_\Gamma=0.55$--$0.63$ of the available rate, below every random-seed measurement ($0.73\pm0.05$; coupling-budget details in End Matter). In log-ratio terms, reduced availability accounts for $60$--$80\%$ of each optimum's growth suppression and reduced alignment accounts for the rest.

The measured thresholds of the optimum's marginals show that the enhancement requires both degrees of freedom. Applying the optimum's phase pattern to uniform amplitudes gives $3.5\,I_\mathrm{mono}$, modestly above the random-phase median of $2.8\,I_\mathrm{mono}$. Its amplitude spectrum with zero phases reaches $4.5\,I_\mathrm{mono}$, comparable to TL at $4.1\,I_\mathrm{mono}$ within the sampled brackets, while the joint design reaches $5.6\,I_\mathrm{mono}$.
Thus, the joint design's advantage over TL resides in the amplitude--phase \emph{correlation} and disappears when either optimized component is removed.
The present budget does not determine how much of the joint advantage arises from temporal waveform correlations and how much from spatial transport, gradients, and multimode dynamics; quantifying their relative contributions is left for future work.

Repeating the joint optimization at ignition-scale conditions ($T_e = 4$~keV, $L_n = 600~\mu$m) yields a threshold that rises as the allowed peak ratio is relaxed before saturating for $R \gtrsim 5$.
Measured against each regime's own coherence-time value, the enhancement is approximately $2.4\times$ in both regimes.
The optimal \emph{design} is regime-specific. Cross-evaluating either regime's optimum in the other delivers only $\approx1.2\times$.
Although both regimes exhibit the same coupling mechanism, each requires a distinct optimal spectrum.
Finding a single spectrum that remains effective across the evolving temperatures, density profiles, and scale lengths of a direct-drive implosion will require optimization over those conditions jointly.
The differentiable solver re-derives each regime-specific spectrum at a cost of roughly one hundred GPU-hours.

Interrupting parametric growth by temporally structuring the drive underlies the STUD-pulse concept~\cite{afeyan_optimal_2013,huller_simulations_2013,albright_control_2014}, which modulates the beam envelope with high on--off contrast to rearrange speckle patterns.
The present approach operates within a single interaction region and a fixed bandwidth budget, tuning spectral amplitudes and phases so that the drive couples weakly, in both magnitude and phase, to the very mode it drives.

Practical implementation depends on both the accessible spectra and the precision of amplitude and phase control.
We have not assessed a specific laser architecture, but its constraints could enter the optimization alongside the peak-ratio penalty.

Coherence time alone does not determine the absolute TPD threshold. Spectral phase and amplitude provide additional control.
At fixed power spectrum, phase optimization raises the threshold but pays for the gain with a TL pulse train of high peak ratio. Joint optimization raises the threshold further while recovering a much lower peak ratio.
The exact pump--field growth-rate budget developed here, $\langle\Gamma_{\mathrm{TPD}}\rangle=\mathcal A_\Gamma\langle\Gamma_{\max}\rangle$, separates the mechanisms. TL acts primarily through temporal concentration and the resulting reduction in normalized time-averaged field magnitude, whereas the joint optima suppress both the available coupling and the fraction realized through pump--daughter alignment. The advantage persists across the simulated $1\%$--$4\%$ bandwidths and under OMEGA- and ignition-scale conditions.
Differentiating through the wave equations allows the search to be repeated for each plasma regime.
Gradient-based inverse design has transformed nanophotonics~\cite{molesky_inverse_2018}. Differentiable plasma simulation~\cite{joglekar_unsupervised_2022,joglekar_gnr_2024} now enables the same search for laser smoothing, whose fielded schemes are single points in the space explored here.
We demonstrate spectral amplitude and phase as control variables for 2-D absolute TPD driven by one prescribed, unspeckled beam.

\begin{acknowledgments}
This material is based upon work supported by IFE COLoR under U.S. Department of Energy Grant No. DE-SC0024863 and by the Department of Energy National Nuclear Security Administration under Award Number DE-NA0004144, the University of Rochester, and the New York State Energy Research and Development Authority. This research used resources of the National Energy Research Scientific Computing Center, a DOE Office of Science User Facility supported by the Office of Science of the U.S. Department of Energy under Contract No. DE-AC02-05CH11231 using NERSC award FES-ERCAP0026741 and AI4Sci@NERSC.
\end{acknowledgments}

\bibliography{references}

\onecolumngrid
\begin{center}
\rule{0.4\linewidth}{0.4pt}\\[0.3em]
{\large\textbf{End Matter}}\\[-0.3em]
\rule{0.4\linewidth}{0.4pt}
\end{center}
\twocolumngrid

\appendix

\begin{figure*}[t]
    \includegraphics[width=\textwidth]{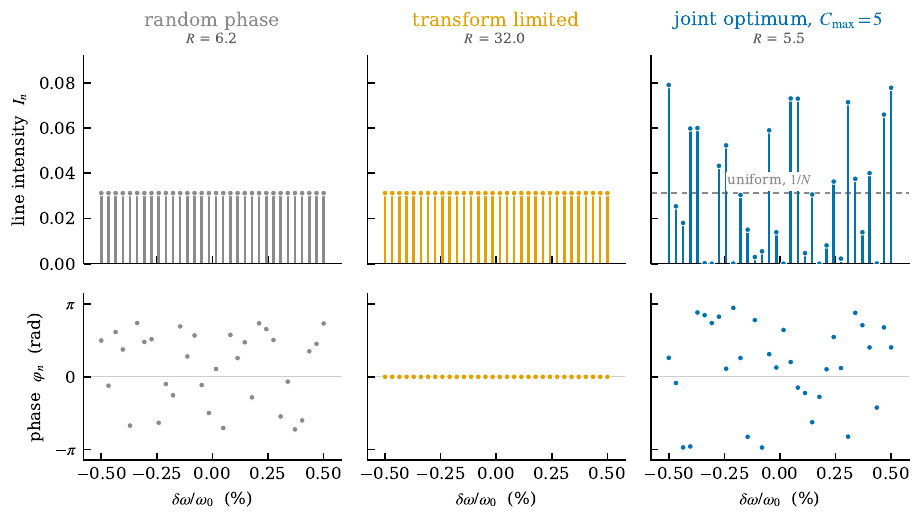}
    \caption{\label{fig:spectra}
    Per-line intensity (top) and spectral phase (bottom) for the $C_\mathrm{max}=5$ joint optimum and two reference drives ($N=32$, $1\%$ bandwidth).
    Random-phase and TL drives share the uniform spectrum $I_n = 1/N$.
    The measured peak-to-average intensity ratio is $R=32.0$ for TL and $5.5$ for the optimum.}
\end{figure*}

\section{Simulation details}\label{app:sim_details}
The 2-D ($p$-polarized) simulations use a single unspeckled beam at normal incidence, absorbing longitudinal boundaries and periodic transverse boundaries, as in the coherence-time calculation~\cite{Follett2019}. The transverse domain is $12~\mu$m; the density range $n_e/n_c\in[0.18,0.28]$ spans $\approx80~\mu$m at OMEGA scale and $240~\mu$m at ignition scale. A continuous, low-pass-filtered stochastic $k$-space source seeds the EPW field ($10^{-10}$ normalized potential units, $k=0$ removed).
The $N$ frequencies uniformly span $[-\Delta\omega/2,+\Delta\omega/2]$; $1\%$ bandwidth means $\Delta\omega=0.01\,\omega_0$.
The grid uses $\Delta x\approx34.1$~nm and $\Delta t\approx1.85$~fs. A resolution sweep and re-optimization on a perturbed grid leave the optima and thresholds unchanged.

\section{Threshold definition and validation}\label{app:protocol}
The threshold observable is $\mathcal E_h(t)=\int |E_h(\mathbf x,t)|^2\,d\mathbf x$. A $40$~ps trajectory is unstable if, during $10$--$40$~ps, $\mathcal E_h$ becomes nonfinite, its endpoint exceeds $10^{-10}$, its duration-normalized temporal mean exceeds $10^{-14}$, or an ordinary-least-squares fit of $\ln\mathcal E_h$ has slope greater than $0.05$~ps$^{-1}$. The $100$~ps tests use $10$--$100$~ps. Threshold is the geometric mean of adjacent stable and sustained-unstable sampled intensities; all higher sampled intensities must also be unstable. The random reference in Fig.~\ref{fig:headline} comprises eight fixed-phase realizations.

The measured monochromatic onset is $1.05\,I_\mathrm{mono}$, within $5\%$ of the analytic threshold~\cite{simon_inhomogeneous_1983}; the two-line onset is $2.07$ times the measured mono onset, consistent with independent beams at half power. Of $90$ trajectories tested to $100$~ps, $86$ retain their $40$~ps classification; the four changes are delayed unstable onsets.

Figure~\ref{fig:bandwidth} samples eight $N=2000$ realizations at $1\%$ and $32$ at each of $2$--$4\%$ bandwidth. The displayed subset illustrates thresholds on both sides of $I_c$, each bracketed within $0.1\,I_c$; TL and joint designs retain $N=32$.

\section{Optimization}\label{app:ad}
The ADEPT solver~\cite{joglekar_adept_2023}, written in JAX~\cite{jax_github_2018}, supplies gradients of
\begin{equation}\label{eq:loss}
\begin{aligned}
\min_{\theta}\;\mathcal{L}(\theta) ={}& \frac{1}{K}\sum_{k=1}^{K}
    \log_{10}\overline{\mathcal{E}}_h(\theta,I_k) \\
    &+\lambda\!\left[\operatorname{relu}\!\left(\widetilde R(\theta)-C_\mathrm{max}\right)\right]^2,
\end{aligned}
\end{equation}
where $\overline{\mathcal E}_h$ is the EPW energy averaged over the final $5$~ps of the phase-only ($40$~ps) or joint ($25$~ps) evolution. Phase-only searches vary $\theta=\{\varphi_j\}$ at fixed equal amplitudes, with $K=1$ and $\lambda=0$. Joint searches vary $\theta=\{A_j,\varphi_j\}$ at fixed $\sum_j A_j^2$, using $K=16$ intensities spanning $[0.7,3.0]\,I_c$. The differentiable log-sum-exp approximation $\widetilde R$ penalizes peak ratio above $C_\mathrm{max}$; $\lambda=1$ for capped designs and $0$ otherwise.
Reverse-mode gradients propagate through the full evolution. Thresholds are measured separately using the protocol above. The optima reach threshold plateaus within the sampled brackets and give lower bounds on the maximum attainable thresholds.

\section{Pump--field coupling budget}\label{app:eta}
We record the complex pump-to-EPW power
\begin{equation}
  \mathcal P_{\mathrm{TPD}}(t)=\int E_h^*\cdot S^{\mathrm{TPD}}\,d\mathbf x,
  \label{eq:Ptpd}
\end{equation}
using the source term and operators of Eq.~\eqref{eq:epw}.
Then $\Gamma_{\mathrm{TPD}}=2\,\mathrm{Re}\,\mathcal P_{\mathrm{TPD}}/\mathcal E_h$ and $\Gamma_{\max}=2|\mathcal P_{\mathrm{TPD}}|/\mathcal E_h$. Linearity of the source in $E_0$ makes $\Gamma_{\max}$ the rate attainable by an instantaneous global pump-phase rotation. The energy budget is $d\ln\mathcal E_h/dt=\Gamma_{\mathrm{TPD}}-\Gamma_{\mathrm{lin}}$, where $\Gamma_{\mathrm{lin}}$ includes damping, boundary loss and longitudinal-projection dissipation.

Over a repetition period $T_{\mathrm{rep}}=2\pi/\Omega_1$, with line spacing $\Omega_1$, the budget-weighted alignment is
\begin{equation}
  \mathcal A_\Gamma
  \equiv \frac{\int_{T_{\mathrm{rep}}}\Gamma_{\mathrm{TPD}}\,dt}
           {\int_{T_{\mathrm{rep}}}\Gamma_{\max}\,dt}
  \;\in\;[-1,1].
  \label{eq:alignment}
\end{equation}
This is the $\Gamma_{\max}$-weighted average of $A(t)=\cos[\arg\mathcal P_{\mathrm{TPD}}(t)]$, so Eq.~\eqref{eq:budget} holds identically. This weighting avoids the exponential-envelope sensitivity of energy-weighted averages.

Figure~\ref{fig:mechanism} uses a common intensity of $1.98\times10^{15}\,\mathrm{W/cm^2}$ ($4.29\,I_c$). Intervals are anchored at each run's noise-floor crossing. TL and optimum points use post-transient periods; random and monochromatic points use the first valid interval because they overflow before a second completes. The monochromatic average spans $0.67$--$4.29$~ps and includes mode establishment. The random reference band combines eleven measurements over two nearby intensities; its mean alignment changes by less than $0.01$ between them.

\end{document}